\documentstyle[twocolumn,aps]{revtex}
\begin{document}
\draft
\preprint{}
\title{Properties of Entanglement Monotones for Three-Qubit Pure States}
\author{R. M. Gingrich}
\address{California Institute of Technology}
\date{\today}
\maketitle
\begin{abstract}
Various parameterizations for the orbits under local unitary
transformations of three-qubit pure states
are analyzed.  The interconvertibility, symmetry
properties, parameter ranges, calculability and behavior under
measurement are looked at.  It is shown that the
entanglement monotones of any multipartite pure state uniquely
determine the orbit of that state under local unitary
transformations.  It follows that there must be an entanglement
monotone for three-qubit pure states which depends on the Kempe
invariant defined in \cite{kempe}.  A form for such an entanglement
monotone is proposed.  A theorem is proved that significantly reduces
the number of entanglement monotones that must be looked at to find
the maximal probability of transforming one multipartite state to
another.  
\end{abstract}
\pacs{}

\newcommand{\sstate}{| \psi \rangle}
\newcommand{\pstate}{| \phi \rangle}

\narrowtext

\section{Introduction}
\label{intro}

Entanglement is at the heart of the studies of quantum computation and
quantum information theory.  It is what separates these studies from
their classical counterparts.  If we are to understand what new
phenomena occur when we look at the true quantum mechanical
description of nature as opposed to the approximations of classical
mechanics then we must understand how the quantum mechanical
description differs from the classical description.  Entanglement is a
measure of this difference.  While entanglement between two parties is
quite well understood \cite{mixedent} \cite{boundent} \cite{nielsen}
\cite{2qminset}, the entanglement within a quantum
algorithm or in a state shared between many parties involves
multipartite entanglement which is just beginning to be understood
\cite{tangle} \cite{eaem} \cite{ntangle}.  

An integral part of the study of entanglement is
determining the  probability of transforming one pure state into another by
Local Operations and Classical Communication (LOCC).  For two part
systems this problem is solved, or at least reduced to the problem of
finding the eigenvalues of a hermitian matrix, by \cite{nielsen}
\cite{2qminset}.  For a $N \times M$ pure state the Schmidt
decomposition tells us we can write 
\begin{equation}
\label{schmidtdecomp}
  \sstate = \sum_{i = 1}^{n} \sqrt{\lambda_{i}^{\uparrow}} | i \rangle
  | i' \rangle
\end{equation}
where the $\lambda_{i}^{\uparrow}$ are in increasing order, $\sum_i \lambda_{i}^{\uparrow} = 1$, the $|i \rangle$ and $|i'
\rangle$ are an orthonormal set of vectors in space $A$ and $B$
respectively,  and
$n = \min(N,M)$.  
If we define 
\begin{equation}
\label{nielsenmon}
E_{k} \left( \sstate \right) = \sum_{i = 1}^{k} 
\lambda_{i}^{\uparrow} \; \; \; \; k = 1, 
\ldots, n - 1  
\end{equation}
then the highest attainable
probability of transforming $\sstate$ to
$\pstate$, $P(\sstate \rightarrow \pstate)$,  is given
by \cite{2qminset}
\begin{equation}
P(\sstate \rightarrow \pstate) = \min\limits_{k} \frac{E_{k} \left(
    \sstate \right)}{E_{k} \left( \pstate \right)}  
\end{equation}
The proof of this theorem is constructive so we
can actually write down the transformation that gives us $\pstate$
from $\sstate$.  For pure states of more than two parts no such nice
theorem is known.  The question of whether two
three-qubit pure states can be transformed into each other with non-zero
probability by LOCC has been solved by D\"ur et.\ al.\ \cite{wghz} but
just getting a
reasonable upper bound on
that probability 
when it is a non-zero is unsolved.  In this paper I
attempt to make some
progress towards solving this problem for three-qubit pure states and
hopefully shed some light on how we might solve it for larger
dimensional spaces and more parts.  

One way to find  $P(\sstate \rightarrow
\pstate)$ is to look at the entanglement monotones $E( \sstate )$ for the two
states.  For the duration of the paper ``state'' will refer to a pure
state unless explicitly called a mixed state.  An entanglement
monotone, EM, is defined as a function that goes
from states to positive real numbers and does not increase under LOCC.
As a convention the
value of any EM for a separable state is 0.  For mixed and pure states
of any dimension and
number of parts the following theorem holds \cite{entmon}
\begin{equation}
\label{probmon}
P(\rho \rightarrow \rho') = \min\limits_{E} \frac{E \left(
    \rho \right)}{E \left( \rho' \right)}
\end{equation}
where the minimization is taken over the set of all EMs
\cite{entmon}.  This can be seen by considering $P ( \rho \to \rho' )$
as an EM for $\rho$.  
The problem is that this minimization is difficult to take
since there is no known way to characterize all the entanglement
monotones for multipartite states.  We would like a ``minimal set'' of
EMs similar to the $E_{k}$
for the bipartite case in order to take the minimization.  

The situation for three or more parts is somewhat different than for
bipartite pure states.  Firstly, generic $M \times M$ bipartite states have a
stabilizer (i.\ e.\ the set of unitaries that takes
a state to itself) of dimension $M - 1$ isomorphic to $U(1)^{\otimes M
  - 1}$ while pure states with more
parts generically have a discrete stabilizer.  States
whose parts are not of the same dimension may have larger stabilizers
but bipartite states are the only ones that always have a continuous
stabilizer.  Secondly, the generalized Schmidt decomposition, however
you choose to generalize it \cite{genschmidt} \cite{tacjal}, has complex
coefficients for pure states
with 3 or more parts.  This implies that generically these states are not
local unitarily equivalent to their complex conjugate
states  (i.\ e.\ the state with each of its coefficients complex
conjugated).  Also, for bipartite pure states all the local unitary (LU)
invariants can be calculated from the eigenvalues of the reduced
density matrices but this does not hold for more parts.  I will go
into more detail about LU invariants in the next section.  

The structure of the paper is as follows, in section \ref{decomps} the interconvertibility, behavior under
measurement, symmetry properties, parameter ranges and calculability
of two generalizations of the Schmidt decomposition of equation
(\ref{schmidtdecomp}) and the polynomial invariants (defined below) are
looked at.  In
section \ref{5thEM} it is shown that the entanglement monotones
uniquely determine the orbit of multipartite pure states and this is
used to show that there must be an EM algebraically
independent of the known EMs.  A form for this EM is proposed and
studied.  Section \ref{OEM}
discusses other monotones that must exist and their properties.
Lastly, in section \ref{minset} a theorem is
proved that significantly reduces the number of EMs that must be
minimized over to get $P(\rho \to \rho')$ of equation (\ref{probmon}).  

\section{Decompositions and Invariants of Three-Qubit Pure States}
\label{decomps}

Let $\sstate$ be a multipartite state in ${\mathcal{H}}_{1} \otimes
{\mathcal{H}}_{2} \ldots \otimes {\mathcal{H}}_{n}$ and let
$A_{k}^{(i)} : {\mathcal{H}}_{i} \rightarrow
{\mathcal{H}}_{i}^{\prime} $ be Krauss operators for an operation on
the hilbert space ${\mathcal{H}}_{i}$ with $\sum_{k}
A_{k}^{(i) \, \dagger}
A_{k}^{(i)} = {\mathcal{I}}_{i}$ and  ${\mathcal{I}}_i$ is the
identity acting on ${\mathcal{H}}_i$.  A (non-increasing) EM is a real valued function
$E \left( \sstate \right) $ such that
\begin{equation}
\label{moninequality}
E \left( \sstate \right) \ge \sum_{k} p_k E \left( \frac{I_{1} \otimes
  \ldots \otimes A_{k}^{(i)} \otimes \ldots \otimes I_{n} \sstate}{\sqrt{p_{k}}} \right)  
\end{equation}
for any state $\sstate$, operation $A_{k}^{(i)}$, and space $i$
where 
\begin{equation}
p_k = \| I_{1} \otimes
  \ldots \otimes A_{k}^{(i)} \otimes \ldots \otimes I_{n} \sstate \|^2 .
\end{equation}
This definition for pure states is taken from the definition for a
general state in \cite{entmon}.  One can always transform a state into product states and a product
state cannot be transformed into anything but another product state so
the value of an EM for a product state is chosen to be zero and all
other states must have a non-negative value for the EM.  Since
$A_{k}^{(i)}$ can be a unitary operator or the inverse of that
operator, equation (\ref{moninequality}) implies that all EMs must be
invariant under LU.  Hence, a first step to understanding the EMs is
to look at the LU invariants that parameterize
the set of orbits.  

There are many ways to find LU invariants for three-qubit states \cite{polyinv} \cite{tacjal}
\cite{todd} \cite{genschmidt} \cite{tangle} \cite{3qinvariants}
\cite{monotones} some of which can be generalized to more parts and
larger spaces but for now I will concentrate on the three-qubit case.
The three sets of invariants I will look at in this section are the polynomial invariants
\cite{polyinv}, what I will call the diagonalization decomposition
\cite{tacjal} and
what I will call the maximization decomposition \cite{genschmidt}.  

\subsection{The Polynomial Invariants}

A general polynomial invariant $P_{\sigma,\tau} \left( \sstate
\right)$ for a state of the form 
\begin{equation}
\sstate = \sum_{i,j,k = 0}^{1} t_{i j k} | i j k \rangle
\end{equation}
is written as
\begin{equation}
\label{polydef}
P_{\sigma,\tau} \left( \sstate \right) = \sum  t_{i_1 j_1 k_1} \ldots
t_{i_n j_n k_n} \bar{t}_{i_1 j_{\sigma
      (1)} k_{\tau (1)}} \ldots \bar{t}_{i_n j_{\sigma (n)} k_{\tau (n)}}
\end{equation}
where $\sigma$ and $\tau$ are permutations on $n$ elements, repeated
indices are summed and $\bar{t}$ stands for the complex conjugate of
$t$ \cite{polyinv}.  If one applies a unitary to any of the qubits in
$\sstate$ and
explicitly writes out $P_{\sigma,\tau} \left( \sstate
\right)$ again it becomes apparent that $P_{\sigma,\tau} \left( \sstate
\right)$ is invariant.  Of course, any polynomial in terms of the
polynomial invariants $P_{\sigma,\tau}
\left( \sstate \right)$ is another polynomial invariant.  In fact, it
can be shown that all the polynomial invariants are of this form.  

We know from \cite{genschmidt} that generic three-qubit states have a discrete
stabilizer so the number of independent polynomial invariants is
given by 
\begin{equation}
\dim \left[ {\mathcal{C}}^2 \otimes {\mathcal{C}}^2 \otimes
{\mathcal{C}}^2 \right] - 3 \dim [SU(2)] - \dim [U(1)] - 1 \nonumber\\
 =  5 
\end{equation}
where the last $- 1$ is due to the fact that we are using
normalized states.
The 5 independent continuous invariants are
\begin{eqnarray}
  I_1 & = & P_{e,(1 2)} \nonumber \\
  I_2 & = & P_{(1 2),e} \nonumber \\
  I_3 & = & P_{(1 2),(1 2)} \nonumber \\
  I_4 & = & P_{(1 2 3),(1 3 2)} \nonumber \\
  I_5 & = & | \sum  t_{i_1 j_1 k_1} t_{i_2 j_2 k_2} t_{i_3 j_3 k_3}
  t_{i_4 j_4 k_4}  \nonumber \\
  & &  \times \epsilon_{i_1 i_2} \epsilon_{i_3 i_4}
  \epsilon_{j_1 j_2} \epsilon_{j_3 j_4} \epsilon_{k_1 i_3}
  \epsilon_{k_2 i_4} |^2
\end{eqnarray}
where $\epsilon_{0 0} = \epsilon_{1 1} = 0$, and $\epsilon_{0 1} = -
\epsilon_{1 0} = 1$ and again repeated indices are summed.  $I_4$ is
the Kempe invariant referred to in the abstract.  If one
writes out $I_5$ and uses the identity $\epsilon_{i j} \epsilon_{r s}
= \delta_{i r} \delta_{j s} - \delta_{i s} \delta_{j r}$ it can be
shown that $I_5$ is just the sum and difference of 64 polynomials of
the form in equation (\ref{polydef}).  With one more discrete invariant, 
\begin{equation}
I_6 = \mbox{sign}\!\! \left[ \mbox{Im}\!\! \left[ P_{(3 4) (5 6),(1 3 5 2 4)}
  \right] \right], 
\end{equation}
the LU orbit of a three-qubit state is determined uniquely \cite{tacjal}
\cite{tacjal2}.  In this paper I will define $\mbox{sign}[x]$ as $1$ for
non-negative numbers and $-1$ otherwise.  The polynomial invariants have
the advantage of being
easy to compute for any state and the four previously known
independent EMs \cite{tangle} are the following simple functions of
$I_1$, $I_2$, $I_3$ and $I_5$
\begin{eqnarray}
\label{monsinI}
\tau_{(AB)C} & = & 2 (1 - I_1) \nonumber \\
\tau_{(AC)B} & = & 2 (1 - I_2) \nonumber \\
\tau_{(BC)A} & = & 2 (1 - I_3) \nonumber \\
\tau_{ABC} & = & 2 \sqrt{I_5}. 
\end{eqnarray}

\subsection{The Diagonalization Decomposition}
\label{DD}

The diagonalization decomposition, DD, introduced by Acin et.\ al.\
\cite{tacjal} is accomplished by first defining matrices $(T_0)_{j,k} = t_{0,j,k}$ and
$(T_1)_{j,k} = t_{1,j,k}$, then finding a unitary operation on space
$A$ that makes $T_0$ singular, finding unitaries on space B
and C that make $T_0$ diagonal and using the remaining phase freedom
to get rid of as many phases as possible.  What is left is a state of
the form
\begin{eqnarray}
| \psi_{\mbox{DD}} \rangle & = & \sqrt{\mu_0}\, | 0 0 0 \rangle +
\sqrt{\mu_1}\, e^{i \phi} | 1 0 0
\rangle \nonumber \\ && + \sqrt{\mu_2}\, | 1 0 1 \rangle +
\sqrt{\mu_3}\, | 1 1 0
\rangle + \sqrt{\mu_4}\, | 1 1 1 \rangle   
\end{eqnarray}
where $\mu_i \ge 0$, $\mu_0 + \mu_1 + \mu_2 + \mu_3 + \mu_4 = 1$ and $0
\le \phi \le \pi$.  Note that generically there are two unitaries that
will make $T_0$ singular but it can be shown that only one will lead
to $\phi$ between $0$ and $\pi$.  If there is another solution, with $\phi$
between $\pi$ and $2 \pi$ exclusive, it is referred to as the dual state
of $| \psi_{\mbox{DD}} \rangle$.  Some nice properties of DD are that
there is a
1 to 1 correspondence with the orbits
and there are a set of invertible functions between the parameters of
the decomposition and the set of polynomial invariants given above.
Namely, 
\begin{eqnarray}
\label{polyDD}
I_1 & = & 1 - 2 \mu_0 ( \mu_2 + \mu_4 ) - 2 \Delta \nonumber \\
I_2 & = & 1 - 2 \mu_0 ( \mu_3 + \mu_4 ) - 2 \Delta \nonumber \\
I_3 & = & 1 - 2 \mu_0 ( \mu_2 + \mu_3 + \mu_4 ) \nonumber \\
I_4 & = & 1 - 3 [( \mu_2 + \mu_3) ( \mu_0 - \mu_4 ) + \mu_4 ( 1 -
\mu_4 ) \nonumber \\ && - \mu_2 \mu_3 \mu_0 + ( 1 - \mu_0 ) ( \Delta -
\mu_1 \mu_4)
] \nonumber \\
I_5 & = & 4 \mu_0^2 \mu_4^2 \nonumber \\
I_6 & = & \mbox{sign} [ \sin ( \phi ) \mu_0^2 \sqrt{\mu_1
    \mu_2 \mu_3 \mu_4 } \nonumber \\ & & \times ( \Delta - \mu_4 (1 - 2
\mu_0 + \mu_1) - \mu_2 \mu_3 ) ]  
\end{eqnarray}
where $ \Delta = \mu_1 \mu_4 + \mu_2 \mu_3 - 2 \sqrt{\mu_1 \mu_2 \mu_3 \mu_4}
\cos (\phi) $ and if we define
\begin{eqnarray}
J_1 & = & \frac{1}{4} \left( 1 - I_1 - I_2 + I_3 - 2 \sqrt{I_5}
\right) \nonumber \\
J_2 & = & \frac{1}{4} \left( 1 - I_1 + I_2 - I_3 - 2 \sqrt{I_5}
\right) \nonumber \\
J_3 & = & \frac{1}{4} \left( 1 + I_1 - I_2 - I_3 - 2 \sqrt{I_5}
\right) \nonumber \\
J_4 & = & \sqrt{I_5} \nonumber \\
J_5 & = & \frac{1}{4} \left( \frac{5}{3} - I_1 - I_2 - I_3 +
  \frac{4}{3} I_4 - 2 \sqrt{I_5} \right) 
\end{eqnarray}
then the coefficients are given by
\begin{eqnarray}
\mu_0^\pm & = & \frac{J_4 + J_5 \pm \sqrt{\Upsilon}}{2 (J_1 + J_4)}
\nonumber \\
\mu_i^\pm & = & \frac{J_i}{\mu_0^\pm}, \;\;\;  i = 2,3,4  \nonumber \\
\mu_1^\pm & = & 1 - \mu_0^\pm - \frac{J_2 + J_3 + J_4}{\mu_0^\pm}
\nonumber \\
\cos (\phi^\pm) & = & \frac{\mu_1^\pm \mu_4^\pm + \mu_2^\pm \mu_3^\pm
  - J_1}{2 \sqrt{\mu_1^\pm \mu_2^\pm \mu_3^\pm \mu_4^\pm}} \nonumber \\
\mbox{sign} [ (\sin (\phi^\pm) ] & = & I_6 \: \mbox{sign} [
  \sqrt{\mu_1^\pm \mu_2^\pm
  \mu_3^\pm \mu_4^\pm} [ J_1 - J_2 J_3 \nonumber \\ && - J_4
(J_2 + J_3 + J_4 - 
(\mu_0^\pm)^2)  ] ]
\end{eqnarray}
where $\Upsilon = (J_4 + J_5)^2 - 4 (J_1 + J_4) (J_2 + J_4) (J_3 +
J_4) \ge 0 $.  The $+$ and $-$ solutions for the coefficients
correspond to $| \psi_{\mbox{DD}} \rangle$ and its dual state.  The
inversion of
the equations for $I_i$ was done independently in \cite{tacjal2}.
Note that their definition of $I_4$ is different from the one in this paper.   

Another nice property of the DD is that we can
perform an arbitrary measurement on it in space $A$ and stay in the DD
form.  Since any measurement can be broken into a series of two
outcome measurements \cite{2outcome}, we can look at the two outcome
measurement $A_1$ and $A_2$ where $A_1^\dagger A_1 + A_2^\dagger A_2 =
I$.  Using the singular value decomposition we can write $A_i = U_i D_i
V$ where $V$ does not depend on $i$ because the two positive hermitian
operators $A_1^\dagger A_1$ and $A_2^\dagger A_2$ sum to the identity
and therefore must be
simultaneously diagonalizable.  The diagonal matrices, $D_i$, can be
written as 
\begin{equation}
D_1 = \left[ \begin{array}{cc} x & 0 \\ 0 & y \end{array} \right] , \;\;\; 
D_2 = \left[ \begin{array}{cc} \sqrt{1 - x^2} & 0 \\ 0 & \sqrt{1 - y^2}
  \end{array} \right] 
\end{equation}
where $0 \le x,y \le 1$ \cite{wghz}.  Since we are only concerned with
what orbit
the outcomes are in we may choose the $U_i$ transformation. 
Also, matrices of the form 
\begin{equation}
\left[ \begin{array}{cc} e^{i \psi_1} & 0 \\ 0 & e^{i \psi_2} 
\end{array} \right]  
\end{equation}
where $\psi_1$ and $\psi_2$ are real numbers, commute with the $D_i$ 
matrices so the most general $V$ can be written as 
\begin{equation}
\left[ \begin{array}{cc} 
\alpha & \sqrt{1 - \alpha^2} e^{i \theta} \\
- \sqrt{1 - \alpha^2} e^{- i \theta} & \alpha
\end{array} \right]  
\end{equation}
where $0 \le \alpha \le 1$ and $\theta$ is real.  If we choose 
\begin{eqnarray}
U_1 & = & \frac{1}{\sqrt{\gamma}} \left[
  \begin{array}{cc} y \alpha & - x \sqrt{1 - \alpha^2} e^{i \theta} \\
x \sqrt{1 - \alpha^2} e^{- i \theta} & y \alpha \end{array} \right]
  \nonumber \\ 
\gamma & = & y^2 \alpha^2 + x^2 (1 - \alpha^2)   
\end{eqnarray}
and similarly for $U_2$ with $(x,y)$ replaced with $(\sqrt{1 -
x^2},\sqrt{1 - y^2})$ then in going from $| \psi_{\mbox{DD}} \rangle $
to $A_1 |
\psi_{\mbox{DD}} \rangle $
the DD coefficients undergo the following transformations
\begin{eqnarray}
\mu_0 & \to & \frac{x^2 y^2 \mu_0}{\gamma} \nonumber \\
\mu_1 & \to & \frac{1}{\gamma} \left| e^{- i \theta} (x^2 - y^2)
  \alpha \sqrt{\mu_0 ( 1 - \alpha^2)} + 
  e^{i \phi} \gamma \sqrt{\mu_1}  \right|^2 \nonumber \\
\mu_i & \to & \mu_i \gamma \; \; \; \; \; \; \; \; i = 2,3,4 \nonumber
\\
\phi &\to & \arg \left[ e^{- i \theta} (x^2 - y^2)
  \alpha \sqrt{\mu_0 ( 1 - \alpha^2)} + 
  e^{i \phi} \gamma \sqrt{\mu_1} \right] 
\end{eqnarray}
and again similarly for $A_2 | \psi_{\mbox{DD}} \rangle$.  Things
become more complicated 
when $\phi$ becomes larger than $\pi$ and we have a dual solution.  In
this case we need to transform to the dual state which can be quite
tedious.  It should also be noted that if we want to plug the new form
for the DD coefficients into equations (\ref{polyDD}) then the
normalization must be taken into account.  The normalization will just
be the sum of the new forms for $\mu_0$ through $\mu_4$.  
 
\subsection{The Maximization Decomposition}

The Maximization Decomposition \cite{genschmidt}, MD, has a somewhat
different way of
decomposing the three qubit states.  First we find the states, $| \phi_A
\rangle$, $ | \phi_B \rangle$ and $ | \phi_C \rangle$ each defined up
to an overall phase,  that maximize 
\begin{equation}
g(| \phi_A \rangle,| \phi_B \rangle,| \phi_C \rangle) = \| \langle
\psi | \phi_A \rangle | \phi_B \rangle |
\phi_C \rangle \|^2 
\end{equation}
and apply a unitary such that $| \phi_A \rangle | \phi_B \rangle|
\phi_C \rangle$ becomes $|0 0 0 \rangle$.  Defining $| 1 \rangle $, up
to an overall phase, as the vector perpendicular to $| 0 \rangle $,
then the derivative of $g$ along $| 1 \rangle $ at the point $| 0 0 0 \rangle
$, 
\begin{eqnarray}
&& \lim_{\epsilon \to 0} \frac{g(| 0 \rangle + \epsilon | 1 \rangle,| 0
\rangle, | 0 \rangle) - g(| 0 \rangle,| 0 \rangle, | 0
\rangle)}{\epsilon} \nonumber \\ && = 2 \mbox{Re} \left[ \langle \psi
| 1 0 0 \rangle
\langle 0 0 0 | \psi \rangle \right]
\end{eqnarray}
must be zero because $g(| 0 \rangle,| 0 \rangle, | 0 \rangle )$ is a
maximum.  Since we still have phase freedom in $| 0 \rangle $ and $| 1
\rangle $ this implies that $\langle \psi | 1 0 0 \rangle = 0 $ and
similarly for $ \langle \psi | 0 1 0 \rangle$ and $ \langle \psi | 0 0
1 \rangle $.  Using the remaining phase freedom in the choice of $|0
\rangle$ and $|1 \rangle$ we can eliminate all but one phase leaving
us with
\begin{equation}
| \psi_{\mbox{MD}} \rangle = a e^{i \phi} | 0 0 0 \rangle + b | 0 1 1 \rangle +
  c | 1 0 1 \rangle + d | 1 1 0 \rangle + f | 1 1 1 \rangle  
\end{equation}
where $ a^2 + b^2 + c^2
+ d^2 + f^2 = 1$, $0 \le \phi \le 2 \pi$,  $0 \le a,b,c,d,f$
and $b,c,d,f \le a$.  Note that $g(|0_A \rangle,|0_B
\rangle, |0_C \rangle) = a^2$.  Unfortunately, the parameters as they are given
above are not in 1 to 1 correspondence with the orbits.  While the
decomposition is generically unique, there are choices of the
parameters within the given ranges that are not the result of the
decomposition.  For example, states with $a^2 = \frac{1}{5} + \epsilon
$, $b^2 = c^2 = d^2 = f^2 = \frac{1}{5} - \frac{\epsilon}{4}$ and any
choice of $\phi$ have 
\begin{equation}
g \left( \frac{1}{\sqrt{2}} ( | 0 \rangle + | 1 \rangle ),
  \frac{1}{\sqrt{2}} ( | 0 \rangle + | 1 \rangle ), \frac{1}{\sqrt{2}}
  ( | 0 \rangle + | 1 \rangle ) \right) \ge a^2 
\end{equation}
for $ \epsilon \le 0.014 $.  Hence, these choices of the parameters
are not a result of the decomposition.  The true ranges of the
parameters that would give a 1 to 1 correspondence with the orbits
are as yet unknown.  

A nice property of the MD is that is it symmetric in particle
exchange.  Exchanging the particles is equivalent to exchanging $b$,
$c$ and $d$.  This makes the permutation properties of the polynomial
invariants easier to
see when written in terms of the MD coefficients.  They take the
following form
\begin{eqnarray}
\label{polyMD}
I_1 & = &  1 - 2 \left( (a^2 + d^2) (b^2 + c^2) + a^2 f^2 \right)
\nonumber \\
I_2 & = &  1 - 2 \left( (a^2 + c^2) (b^2 + d^2) + a^2 f^2 \right)
\nonumber \\
I_3 & = &  1 - 2 \left( (a^2 + b^2) (c^2 + d^2) + a^2 f^2 \right)
\nonumber \\
I_4 & = &  1 - 3 ( a^2 (1 - a^2) - ( b^2 c^2 + b^2 d^2 + c^2 d^2 ) (1
- 2 a^2) \nonumber \\ 
&& - 2 b^2 c^2 d^2 - 2 a b c d f^2 \cos (\phi)) \nonumber \\
I_5 & = &  a^2 |\, a f^2 + 4 b c d e^{i \phi}|^2 \nonumber \\
I_6 & = & \mbox{sign} [ a b c d f^2 \sin (\phi) ( a^2 (1 - 2
  a^2)(1 - 2 a^2 - f^2) \nonumber \\ 
&& - 4 b^2 c^2 d^2 - 2 a b c d  f^2 \cos (\phi) ) ].
\end{eqnarray}
It is apparent from these equations that $I_1$, $I_2$ and $I_3$ are
symmetric in permutations of particles $A B$, $A C$ and $B C$
respectively and $I_4$, $I_5$ and $I_6$ are symmetric in any
permutation of the particles.  Unfortunately, the equations in
(\ref{polyMD}) are not as
easy to invert as those in (\ref{polyDD}).  In fact, just calculating the
MD coefficients for an arbitrary state is not an easy task, as it is in
the case of the polynomial invariants and the DD coefficients, since
determining the unitaries for the MD 
involves maximizing over a 6 dimensional space with typically many
local maxima.  

One more interesting fact about the MD is that $1 - a^2$ is a
non-increasing EM.  We know this because in
\cite{monotones} it is shown that a function of the form
\begin{equation}
\label{lbmon}
  E_{k_A,k_B,k_C} (\sstate) = \max\limits_{\Gamma_A,\Gamma_B,\Gamma_C}
  \| \Gamma_A \otimes \Gamma_B \otimes \Gamma_C \sstate \|^2
\end{equation}
where $\Gamma_X$ is a $k_X$-dimensional projector on system $X =
A,B,C$, is an non-decreasing EM and $E_{1,1,1} (\sstate) = a^2$.  The EM
$1 - a^2$ can be shown to be independent of the $\tau$ from equation
(\ref{monsinI}) by looking at the gradient vectors of the $\tau$, $1 -
a^2$ and $N = a^2 + b^2 + c^2 + d^2 + f^2$ at, for instance, the point
$a = 3$, $b,c,d,f =
1$ and $\phi = \frac{\pi}{2}$.  Since the gradient vectors span a 6
dimensional space, $1 - a^2$ cannot be written in terms of the $\tau$
and $N$.  The problem with using $1 - a^2$ as an EM is that one needs
to find the global maximum of a 6 dimensional space with many local
maxima to calculate it.  This is a difficult task for most states.  

\section{fifth independent EM}
\label{5thEM}

In section \ref{decomps} it was shown that all EMs must be invariant
under LU and hence are determined by the orbit of the state.  For three
qubit states this means that EMs are a function of only the
polynomial invariants, DD coefficients or MD coefficients.  In fact,
this determination is unique.  
\newtheorem{theorem}{Theorem}
\label{multiEM}
\begin{theorem} The set of all EMs for any multipartite pure state, $\sstate$,
  uniquely determine the orbit of the state.  
\end{theorem}
{\bfseries Proof.} Suppose two states $\sstate$ and $\pstate$ in
${\mathcal{H}}_{1} \otimes
{\mathcal{H}}_{2} \ldots \otimes {\mathcal{H}}_{n}$  have the
same values for the EMs but lie in different orbits.  We know by using
equation (\ref{probmon}) that  
\begin{equation}
P(\sstate \rightarrow \pstate) = P(\pstate \rightarrow \sstate) = 1
\end{equation} 
so $\sstate$ can be transformed to $\pstate$ (and vice versa) by $n$-party
LOCC, $n$-LOCC, with probability 1.  Since EMs are
non-increasing with any $n$-LOCC they must remain constant during the
entire transformation from $\sstate$ to $\pstate$ (and vice versa).  
Also, we know
that any EM between a system $X = A,B, \ldots$ and
the rest of the
systems thought of as one (e.\ g.\ between $B$ and $(A C D \ldots)$),
I will call these EMs 2-EMs, 
is also
an EM for multipartite states.  This is because any $n$-LOCC on
the multipartite state is also a 2-LOCC between $X$ and the rest
of the systems, since the 2-EM is non-increasing over 2-LOCC it must also
be non-increasing over
$n$-LOCC.  In particular the sum of the lowest $k$ eigenvectors of the
reduced density matrices,
\begin{equation}
E_k^X ( \sstate ) = \sum_{i = 1}^{k} \lambda_{i}^{\uparrow} (\rho_X
(\sstate)), 
\end{equation}
(i.\ e.\ the 2-EMs in equation
(\ref{nielsenmon})) must be EMs.  So the $E_k^X ( \sstate )$ must
remain unchanged and hence the spectrum of $\rho_X$ is unchanged
during the transformation from $\sstate$ to $\pstate$. In
particular a measurement on space
$X$, given by $A_1$ and $A_2$, must be such that 
\begin{equation}
\rho_X \left( \frac{A_i \sstate}{\sqrt{N}} \right) = U \rho_X \left(
  \sstate \right) U^\dagger 
\end{equation}
where $N$ is the normalization.  The only way this can be satisfied is
if $\frac{A_i}{\sqrt{N}}$ is a unitary matrix. This means that $\sstate$ and
$\pstate$ are unitarily equivalent which contradicts our original
supposition.  $\Box$
 
Since we know there are 5 parameters that determine the orbit of a
three qubit state then by theorem 1 there must be 5
independent, continuous EMs.  To the best of the author's knowledge
the only 4 known independent continuous EMs that don't require a
difficult maximization over a multidimensional space are the four
$\tau$ EMs defined in
equation (\ref{monsinI}).  Any candidate for the fifth independent
EM must depend on $I_4$ since the $\tau$ are invertible
functions of $I_1$, $I_2$, $I_3$ and $I_5$ respectively.  The
following function fulfills that criterion
\begin{equation}
\sigma_{ABC} = 3 - (I_1 + I_2 + I_3) I_4  
\end{equation}
and numerical results suggest that it is an EM.  After generating over
300,000 random states and applying a random operation to each of them
the inequality in equation
(\ref{moninequality}) was never violated by $\sigma_{ABC}$.  Also, note
that $\sigma_{ABC}$ is symmetric in particle permutations as is
$\tau_{ABC}$.  For the rest of the paper I will assume that
$\sigma_{ABC}$ is an EM.  Indeed, it may be that there is a set of
measure zero or perhaps just a very small measure for which
$\sigma_{ABC}$ is not a monotone and my
numerical test didn't explore this space but there must exist
some function of the polynomial invariants which is independent of the
$\tau$s and is an EM.  For it to be useful in improving our
upper bound for $P
\left( \sstate \rightarrow \pstate \right)$ there should be pairs of
states $\sstate$
and $\pstate$ such that
\begin{equation}
\frac{\sigma_{ABC}(\sstate) }{\sigma_{ABC}(\pstate)} <
   \min\limits_{\tau}
   \frac{\tau \left(
    \sstate \right)}{\tau \left( \pstate \right)} 
\end{equation}
and I have found such states numerically.    
The largest value of 
\begin{equation}
  \frac{\sigma_{ABC}(\sstate) }{\sigma_{ABC}(\pstate)} -
   \min\limits_{\tau}
   \frac{\tau \left(
    \sstate \right)}{\tau \left( \pstate \right)} 
\end{equation}
that I found in my limited number of examples of was 0.01 and I was
able to find examples of states for which $\tau \left(
    \sstate \right) / \tau \left( \pstate \right)$ is greater than
one for all $\tau$ and $\sigma_{ABC}(\sstate)
/ \sigma_{ABC}(\pstate)$ is less than one.

\section{Other EMs and the Discrete Invariant}
\label{OEM}

The five independent continuous EMs, $\tau_{(AB)C}$, $\tau_{(AC)B}$,
$\tau_{(BC)A}$, $\tau_{ABC}$ and $\sigma_{ABC}$, can easily be
inverted to find $I_1$ - $I_5$ but to completely determine the orbit of
a state we must also have an EM that will give us the value of the
discrete invariant $I_6$.  This is equivalent to finding an EM that is
not the same for a state and it complex conjugate state.  Note that
$I_1, \ldots I_5$ and hence the $\tau$ and $\sigma_{ABC}$ do not
change when a state is conjugated but by looking at any of the sets
of LU invariants we can see that generically a state is not LU
equivalent to its conjugate.  By looking at equation (\ref{probmon}) we
can see that this implies that there must be EMs that are not the
same for the generic state and its conjugate.  It is also easy to see that
for any operation that takes a state $\sstate$ to its conjugate
$\bar{\sstate}$ with probability $p$ there is an operation that takes
$\bar{\sstate}$ to $\sstate$ with the same probability.  So, for a
generic state $\sstate$ there
must be an EM that goes down for the operation $\sstate \to
\bar{\sstate}$ and a similar one that goes down the same amount for
$\bar{\sstate} \to \sstate$.  So, EMs of the following form must
exist 
\begin{equation}
\label{uppm}
\upsilon^{\pm} \left( \sstate \right) = \left\{ \begin{array}{cc}
    \upsilon + \upsilon' & \; \; \; \pm I_6 = 1 \\ \upsilon & \; \; \;
    o.w. 
\end{array} \right.  
\end{equation}
where $\upsilon$ and $\upsilon'$ are functions of $\tau_{(AB)C}$,
$\tau_{(AC)B}$, $\tau_{(BC)A}$, $\tau_{ABC}$ and $\sigma_{ABC}$.  

Also, from \cite{wghz} we know that there are two classes of three-part
entangled states (i.\ e.\ states with $\tau_{(AB)C},\tau_{(AC)B},\tau_{(BC)A}
> 0$) that can be converted into each other with some non-zero
probability within the class and zero probability between the classes.
Namely, the GHZ-class which contains 
\begin{equation}
| \mbox{GHZ} \rangle = \frac{1}{\sqrt{2}} \left( | 000 \rangle + | 111
  \rangle  \right)  
\end{equation}
and has non-zero $\tau_{ABC}$ and the W-class which contains
\begin{equation}
| \mbox{W} \rangle = \frac{1}{\sqrt{3}} \left( | 001 \rangle + | 010
  \rangle + | 100 \rangle \right). 
\end{equation}
and has $\tau_{ABC} = 0$.  Looking again at equation (\ref{probmon})
we see that $\tau_{ABC}$ tells us that $P( | \psi_{\mbox{W}} \rangle \to |
\psi_{\mbox{GHZ}} \rangle) = 0 $ but none of the previously
defined EMs tell us that $P( | \psi_{\mbox{GHZ}} \rangle \to |
\psi_{\mbox{W}} \rangle) = 0 $.  Since the only way to get $P( |
\psi_{\mbox{GHZ}} \rangle \to |
\psi_{\mbox{W}} \rangle) = 0$ is to have an EM that is finite for
GHZ-class states and infinite for W-class states or zero for GHZ-class states
and non-zero for W-class states such an EM must exist.

\section{Finding a Minimal Set}
\label{minset}

Since $\tau_{(AB)C}$, $\tau_{(AC)B}$, $\tau_{(BC)A}$, $\tau_{ABC}$,
$\sigma_{ABC}$ and $\upsilon^\pm$ determine the orbit of the state all
other EMs must depend on them.  A fairly general way to create new EMs from
known  EMs is to use what I will call
$f$-type functions  
\newtheorem{definition}{Definition}
\begin{definition} A function $f : {\mathcal{S}} \subset \Re^n \to \Re$
  is an $f$-type
  function if it satisfies the following
  \begin{enumerate}
  \item $f( \vec{0} ) = 0$
  \item if $x_i \ge y_i$ for all $i = 1,2,\ldots n$ then $f(\vec{x})
    \ge f(\vec{y})$ for $\vec{x}, \vec{y} \; \epsilon \; {\mathcal{S}}$
  \item $f( p \vec{x} + (1 - p) \vec{y}) \ge p f(\vec{x}) + (1 - p)
    f(\vec{y})$ for any $\vec{x}, \vec{y} \; \epsilon \;
    {\mathcal{S}}$ and $0 \le p \le 1$.
  \end{enumerate}
\end{definition}
For a set of EM, $\{ E_i \}$, we have 
\begin{equation}
  E_i ( \sstate ) \ge p E_i \left( \frac{A_1 \sstate}{\sqrt{p}}
  \right) + (1 - p)
  E_i \left( \frac{A_2 \sstate}{\sqrt{1 - p}} \right).
\end{equation}
for any measurement $A_1$, $A_2$ and any state $\sstate$.  So, we have 
\begin{eqnarray}
  f[ \vec{E} (\sstate) ] & \ge & f \left[ p \vec{E} \left( \frac{A_1
  \sstate}{\sqrt{p}}
  \right) + (1 - p)
  \vec{E} \left( \frac{A_2 \sstate}{\sqrt{1 - p}} \right) \right]
  \nonumber \\
  & \ge & p f \left[ \vec{E} \left( \frac{A_1
  \sstate}{\sqrt{p}} \right) \right] + (1 - p) f \left[ \vec{E} \left(
  \frac{A_2 \sstate}{\sqrt{1 - p}} \right) \right] \nonumber \\
  &&
\end{eqnarray}
where the first inequality comes from property 2 and the second comes
from property 3.  Hence, $f(E_1, \ldots ,E_m )$ is also an EM.  
We can show that any EM $f(E_1, \ldots ,E_m)$ that is
an $f$-type function of
monotones $E_1, \ldots E_m$ does not modify the upper bound on
$P(\sstate \to \pstate)$ given by 
\begin{equation}
  P(\sstate \to \pstate) \le \min\limits_{i} \frac{E_{i} \left(
    \sstate \right)}{E_{i} \left( \pstate \right)}.
\end{equation}
First for the one dimensional case.  
\newtheorem{lemma}{Lemma}
\begin{lemma}
If $f(x)$ is an $f$-type function with $n = 1$ then 
\begin{equation}
\frac{f(x)}{f(y)} \ge \min \left\{ \frac{x}{y}, 1 \right\}  
\end{equation}
for any $x,y \; \epsilon \; {\mathcal{S}}$.  \end{lemma}
{\bfseries Proof.} \begin{description}
\item[Case 1] For $x \ge y$ from property 2 we know $f(x) \ge f(y)$
  and hence 
  \begin{equation}
  \frac{f(x)}{f(y)} \ge 1.  
  \end{equation}
\item[Case 2] For $x < y$ if we choose $p = \frac{x}{y} \; \epsilon \; [0,1)$
then we know from properties 1 and 3 that $f(p y) \ge p f(y)$ and so 
\begin{equation}
\frac{f(x)}{f(y)} \ge \frac{x}{y}. \: \Box 
\end{equation}
\end{description}

For $n$ dimensions we have the following theorem (proved with
S. Daftuar and D. Whitehouse).  
\newtheorem{theorem2}[theorem]{Theorem}
\begin{theorem}
If $f(x)$ is an $f$-type function then 
\begin{equation}
  \frac{f(\vec{x})}{f(\vec{y})} \ge \min \left\{ \frac{x_i}{y_i},1
  \right\} \; \; \; \; \; i = 1, 2, \ldots n
\end{equation}
for $\vec{x},\vec{y} \; \epsilon \; {\mathcal{S}}$. 
\end{theorem} 
{\bfseries Proof.} Let 
\begin{equation}
c = \min \left\{ \frac{x_i}{y_i} \right\}  
\end{equation}
then we have
 \begin{description}
\item[Case 1] If $c \ge 1$ then from property 2 $f(\vec{x}) \ge
  f(\vec{y})$ and so
  \begin{equation}
    \frac{f(\vec{x})}{f(\vec{y})} \ge 1
  \end{equation}
\item[Case 2] If $c < 1$ then define 
  \begin{equation}
    z_i = \frac{x_i}{c} \; \; \; \; \; i = 1, 2, \ldots n
  \end{equation}
and $g(r) = f(r \vec{z})$.  Notice that $g(r)$ is an $f$-type function
with $n = 1$ and hence 
\begin{equation}
  \frac{g(c)}{g(1)} \ge c
\end{equation}
or substituting in $f$ we have
\begin{equation}
  \frac{f(\vec{x})}{f(\vec{z})} \ge c.
\end{equation}
Using $z_i \ge y_i$ and property 2 we have
\begin{equation}
  \frac{f(\vec{x})}{f(\vec{y})} \ge c. \: \Box
\end{equation}
\end{description}
For three-qubit states if we take the minimum of $ E(\sstate) /
E(\pstate)$ over ${\mathcal{E}} = \{ \tau_{(AB)C},
\tau_{(AC)B}, \tau_{(BC)A}, \tau_{ABC}, \sigma_{ABC}, \upsilon^\pm \}$ we
are actually taking the minimum over the infinite
set of all $f$-type
functions of ${\mathcal{E}}$.  Although from theorem 1 we know that
all EMs must
be a function of ${\mathcal{E}}$ it is possible that there exist EMs
that are not
$f$-type functions of ${\mathcal{E}}$.  These EMs could cause
$P(\sstate \to \pstate)$ to be lower than the minimum of $E(\sstate) /
E(\pstate)$ over ${\mathcal{E}}$.  The EM
mentioned at the end of section \ref{OEM} is an example of such an EM. 

\section{Conclusions and Further Research}

Theorem 1 along with theorem 2 implies that there should be a (not
necessarily finite) minimal set of EMs, $M$, for which all EMs for
three-qubit states or
similarly for any type of multipartite states are $f$-type functions
of $M$.  I conjecture that such a minimal set should be simple since
the $f$-type functions
seem to be a rather general way of creating EMs that are functions of
other EMs.  The difficult part seems to be finding the EMs that are minimal and
showing that they are minimal.  Using numerical results it seems that
the $\tau$ may be minimal.  I looked at functions of the $\tau$ that
are almost but not quite $f$-type such as $\tau^{1.01}$ and
numerically tested whether they are EMs or not.  None of them were
EMs.  I cannot say the same for $\sigma_{ABC}$ and definitely not for
$\upsilon^\pm$ since I do not have an explicit form for the
$\upsilon$.  

There is further research that may help these problems.  If one could
invert the equations in (\ref{polyMD}) to write $a,b,c,d,f$ and $\phi$ in
terms of $I_1, \ldots , I_6$ that would allow us to calculate the EM
$1 - a^2$ not to mention find the ranges for and
calculate the values of $a,b,c,d,f$ and $\phi$.  The EM $1 - a^2$ could be used
to replace $\sigma_{ABC}$ or perhaps as an addition to
${\mathcal{E}}$ and may prove more useful than $\sigma_{ABC}$.  As far
as finding the minimal EMs and showing that
they are minimal, the arbitrary measurement on the DD at the end of
section (\ref{DD}) may be useful since it allows us to look at the
value of $I_1, \ldots , I_6$ before and after an arbitrary measurement
on an arbitrary state with far less parameters than if we didn't take
out the LU freedom.  Also, it may be able to tell us the maximal
probability of transforming the general complex state $\sstate$ to its
conjugate state $| \bar{\psi} \rangle$ and this is a crucial piece
of information that is needed to calculate $\upsilon'$ in equation
(\ref{uppm}).  Unfortunately, most of these tasks involve trying to
solve nontrivial equations or systems of equations with many variables
which can be difficult or even impossible.

\section{Acknowledgments}
I would like to thank my Advisor John Preskill for supporting me
during this research and for many helpful discussions.  I would also
like to thank Todd Brun, Sumit Daftuar, Julia Kempe, Michael Nielsen, Federico
Spedalieri, Frank Verstraete, Guifre Vidal, Anthony Sudbery and David
Whitehouse for interesting discussions.

\bibliographystyle{prsty}
\bibliography{entbib}

\end{document}